\documentclass{article}

\usepackage{PRIMEarxiv}

\usepackage[utf8]{inputenc} 
\usepackage[T1]{fontenc}    
\usepackage{hyperref}       
\usepackage{url}            
\usepackage{booktabs}       
\usepackage{amsfonts}       
\usepackage{microtype}      
\usepackage{fancyhdr}       
\usepackage{graphicx}       

\usepackage{amsmath,amssymb}
\usepackage{longtable,array,multirow}
\usepackage{float}
\usepackage{colortbl}
\usepackage{pdflscape}
\usepackage{tabularx}
\usepackage{xltabular}
\usepackage{threeparttable}
\usepackage{threeparttablex}
\usepackage[normalem]{ulem}
\usepackage{makecell}
\usepackage{xcolor}

\makeatletter
\newsavebox\pandoc@box
\newcommand*\pandocbounded[1]{
  \sbox\pandoc@box{#1}%
  \Gscale@div\@tempa{\textheight}{\dimexpr\ht\pandoc@box+\dp\pandoc@box\relax}%
  \Gscale@div\@tempb{\linewidth}{\wd\pandoc@box}%
  \ifdim\@tempb\p@<\@tempa\p@\let\@tempa\@tempb\fi
  \ifdim\@tempa\p@<\p@\scalebox{\@tempa}{\usebox\pandoc@box}%
  \else\usebox{\pandoc@box}%
  \fi%
}
\def\fps@figure{htbp}
\makeatother

\NewDocumentCommand\citeproctext{}{}
\NewDocumentCommand\citeproc{mm}{%
  \begingroup\def\citeproctext{#2}\cite{#1}\endgroup}
\makeatletter
 \let\@cite@ofmt\@firstofone
 \def\@biblabel#1{}
 \def\@cite#1#2{{#1\if@tempswa , #2\fi}}
\makeatother
\newlength{\cslhangindent}
\newlength{\csllabelwidth}
\newenvironment{CSLReferences}[2] 
 {\begin{list}{}{%
  \setlength{\itemindent}{0pt}
  \setlength{\leftmargin}{0pt}
  \setlength{\parsep}{0pt}
  \ifodd #1
   \setlength{\leftmargin}{\cslhangindent}
   \setlength{\itemindent}{-1\cslhangindent}
  \fi
  \setlength{\itemsep}{#2\baselineskip}}}
 {\end{list}}
\usepackage{calc}

\providecommand{\tightlist}{%
  \setlength{\itemsep}{0pt}\setlength{\parskip}{0pt}}

\title{Bayesian epidemic alignment for causal evaluation of seasonal infectious-disease interventions}

\author{
  David Moriña \\
  Departament de Matemàtiques, \\
  Universitat Autònoma de Barcelona, \\
  Cerdanyola del Vallès, Spain \\
  \texttt{David.Morina@uab.cat} \\
}

\begin{document}
\maketitle

\begin{abstract}
Seasonal infectious-disease interventions are commonly evaluated with interrupted time-series or pre--post designs that align epidemics by calendar week. When epidemic onset, speed or peak timing differs between seasons, such comparisons confound a shift in epidemic phase with a change in disease burden. A Bayesian causal count model is proposed in which season-specific affine transformations map calendar time to a latent epidemic clock, and intervention effects are estimated on that clock rather than on the calendar. The alignment is a model component rather than a preprocessing step, so uncertainty about epidemic timing propagates into every causal contrast. The model uses a negative-binomial observation distribution, hierarchical area, season and area-season effects, a shrunk Fourier epidemic curve, and a continuous programme-intensity exposure. Posterior g-computation yields prevented cases, prevented fractions, peak attenuation and epidemic displacement, under both a controlled contrast and a dynamic contrast that propagates disease history within each arm. A two-tier simulation study evaluates bias, root mean squared error, interval coverage and parameter recovery under stable timing, epidemic-clock variation, intensity-dependent ascertainment and area-level confounding. The proposed framework is illustrated using open Catalan primary-care surveillance and respiratory syncytial virus immunisation data, with explicit attention to the overlap in programme intensity that identifies the effect.

\end{abstract}


\section{Introduction}\label{introduction}

Population-level interventions against seasonal infectious diseases are
usually evaluated with interrupted time-series designs,
difference-in-differences estimators or Bayesian structural time-series
models. Each expresses the intervention effect as a departure from a
counterfactual trajectory, and each can accommodate temporal dependence
and contemporaneous controls (\citeproc{ref-brodersen2015}{Brodersen et
al. 2015}). Almost all of them index seasonality by calendar week.

For respiratory infections that convention is consequential. Onset,
growth rate, peak and decay routinely move by several weeks between
seasons, for reasons ranging from meteorological variation to
displacement by other circulating pathogens. A post-intervention
epidemic that peaks two weeks late will, under calendar alignment,
register as a burden reduction in the weeks around the historical peak.
A genuinely effective programme in a season that starts early will
register as weaker than it is. The error does not average out: it has
the sign of the timing change.

Functional data analysis separates phase variation, the
horizontal displacement or deformation of a curve, from amplitude
variation, its vertical magnitude (\citeproc{ref-ramsay1998}{Ramsay and
Li 1998}). Deterministic curve registration estimates the phase first
and analyses the aligned curves afterwards, which discards the
uncertainty in the alignment. Bayesian registration instead treats the
temporal deformation as a latent quantity and propagates its uncertainty
(\citeproc{ref-telesca2008}{Telesca and Inoue 2008};
\citeproc{ref-kim2022}{Kim, Chkrebtii, and Kurtek 2022};
\citeproc{ref-gardella2025}{Gardella et al. 2025}). That machinery has
not been carried into causal evaluation of seasonal interventions, where
the estimand of interest is a change in amplitude and the nuisance is
precisely a change in phase.

This work describes a Bayesian causal count model in which the phase--amplitude
separation is internal to the estimation. Season-specific affine maps
carry calendar time to a latent epidemic clock; the epidemic curve, the
hierarchical structure and the intervention effect are all defined on
that clock; and posterior g-computation returns counterfactual
trajectories under a zero-intensity programme. The observation model
accommodates the features that surveillance data require:
overdispersion, geographical heterogeneity, syndromic rather than
pathogen-specific diagnoses, and short-term dependence
(\citeproc{ref-held2006}{Held et al. 2006};
\citeproc{ref-paul2008}{Paul, Held, and Toschke 2008};
\citeproc{ref-corberan2014}{Corberán-Vallet and Lawson 2014}).

The motivating application is the introduction of Nirsevimab against
respiratory syncytial virus (RSV) in Catalonia, where daily primary-care
syndromic surveillance and campaign-level immunisation counts are
published as open data, and where substantial
reductions in bronchiolitis and severe RSV outcomes have been reported
(\citeproc{ref-coma2024}{Coma et al. 2024},
\citeproc{ref-coma2025}{2025};
\citeproc{ref-perramon2024}{Perramon-Malavez et al. 2024}). The
application also illustrates a point that generalises beyond it: a
continuous-intensity exposure identifies an effect only to the extent
that intensity varies across areas within a programme season, and that
overlap must be established before any effect estimate is interpreted.

This article (i) presents an epidemic-aligned causal model for overdispersed
surveillance counts and states its identification conditions; (ii)
evaluates its operating characteristics under phase variation, differential 
ascertainment and area-level confounding; and (iii) applies
it to Catalan open data.

\section{Methods}\label{methods}

\subsection{Notation, potential outcomes and
estimands}\label{notation-potential-outcomes-and-estimands}

Let \(Y_{rst}\) be the number of syndromic events in basic health area
\(r\) (\emph{àrea bàsica de salut}, ABS; see Appendix A), epidemic
season \(s\) and week \(t\), with population at risk \(N_{rst}>0\). Let
\(I_{rs}\geq 0\) denote scaled programme intensity and
\(T_s=\mathbb{I}(s\geq s_0)\) indicate that the programme was available.
In the application, \(Y_{rst}\) counts weekly primary-care bronchiolitis
diagnoses among infants and \(I_{rs}\) measures area-specific Nirsevimab
activity, scaled so that zero denotes no recorded activity and one the
median intensity among programme ABS--seasons. An ABS--season is one
basic health area observed over one July--June season, and is the unit
at which the exposure varies.

Write \(Y_{rst}(i)\) for the potential outcome under intensity \(i\).
The two primary estimands are the prevented count and prevented
fraction,

\[
C_{rs}=\sum_t\bigl\{Y_{rst}(0)-Y_{rst}(I_{rs})\bigr\},
\qquad
PF_{rs}=1-\frac{\sum_tY_{rst}(I_{rs})}{\sum_tY_{rst}(0)} .
\]

The same posterior counterfactual trajectories yield peak attenuation,
displacement of the week of maximum incidence, time above an incidence
threshold, and contrasts across geographical or socioeconomic strata.
Identification requires consistency, positivity over the intensity
contrasts being evaluated, and conditional exchangeability given disease
history, area, season and area-season effects, socioeconomic composition
and the epidemic clock. Positivity is the binding condition in practice,
which is further discussed in Section 3.2.

\subsection{The epidemic clock}\label{the-epidemic-clock}

Calendar week \(t\) in season \(s\) maps to latent epidemic time by the
affine transformation

\[
\tau_s(t)=\{t-t_0-\delta_s\}\exp(\kappa_s),
\]

with shift \(\delta_s\) in weeks, log-speed \(\kappa_s\), and a fixed
pivot \(t_0\) common to all seasons, taken as the mean week of the
analysis window. The map is strictly increasing, so temporal order is
preserved. Positive \(\delta_s\) places a given epidemic phase later in
the calendar; \(\exp(\kappa_s)>1\) compresses the season and
\(\exp(\kappa_s)<1\) stretches it. If the common epidemic curve peaks at
latent time \(\tau^\star\), the implied calendar peak in season \(s\) is
\(t^\star_s=t_0+\delta_s+\tau^\star\exp(-\kappa_s)\), which separates a
shift in onset from a change in tempo. The pivot carries no substantive
content, but placing it inside the observation window rather than at
week zero, which lies a full half-season outside it, substantially
reduces the posterior dependence between \(\delta_s\) and \(\kappa_s\)
and improves sampling geometry.

A common displacement of all \(\delta_s\), or a common rescaling of all
\(\kappa_s\), is absorbable into the phase and frequency of the baseline
curve. The clock must therefore be anchored. Exact zero-sum constraints 
are imposed on the transformed parameters,

\[
\sum_{s=1}^S\delta_s=0,\qquad \sum_{s=1}^S\kappa_s=0,
\]

so that the latent clock is defined relative to the average season,
using the isometric log-ratio (Helmert) basis: for
\(\boldsymbol{z}\in\mathbb{R}^{S-1}\) with
\(z_j\sim N\{0,(1-S^{-1})^{-1}\}\), the mapped vector has exactly zero
sum, unit marginal variances and exchangeable correlation
\(-(S-1)^{-1}\). Writing
\(\delta_s=\sigma_\delta\,\tilde{z}_{\delta s}\) and
\(\kappa_s=\sigma_\kappa\,\tilde{z}_{\kappa s}\) for mapped vectors
\(\tilde{\boldsymbol{z}}\), the scale parameters \(\sigma_\delta\) and
\(\sigma_\kappa\) are directly interpretable as the between-season
standard deviations of epidemic onset and log speed. As
\(\sigma_\delta,\sigma_\kappa\) go to 0 the model reduces to an ordinary
calendar-time seasonal regression, so calendar alignment is a special
case rather than an alternative.

\subsection{Observation model}\label{observation-model}

Counts follow a negative binomial in the mean--dispersion
parameterisation,
\(Y_{rst}\mid\mu_{rst},\phi\sim\mathrm{NB}_2(\mu_{rst},\phi)\), with
\(\mathrm{Var}(Y)=\mu+\mu^2/\phi\). The conditional mean is

\[
\begin{aligned}
\log\mu_{rst}={}&\log N_{rst}+\alpha+u_r+v_s+w_{rs}+f\{\tau_s(t)\}\\
&+\beta_LL_{rst}+\beta_{\mathrm{SES}}Z_{rs}+T_sI_{rs}\,m\{\tau_s(t)\},
\end{aligned}
\]

where \(L_{rst}\) is the standardised previous-week value of
\(\log(1+Y)\) and \(Z_{rs}\) the standardised socioeconomic index. The
population offset places all coefficients on the incidence scale.

The hierarchical terms are \(u_r\sim N(0,\sigma_u^2)\) for persistent
area risk, \(v_s\sim N(0,\sigma_v^2)\) for season amplitude not
explained by timing, and \(w_{rs}\sim N(0,\sigma_w^2)\) for area--season
departures. The exposure \(I_{rs}\)
is constant across the weeks of an ABS--season, so without \(w_{rs}\)
the \(52\) weekly observations in a block enter the likelihood as
independent replicates of a single exposure contrast. The effective
sample size for the intervention coefficient is the number of
ABS--seasons, not the number of weeks, and omitting \(w_{rs}\) produces
credible intervals that are too narrow by a factor that grows with the
within-block correlation. Exchangeable area effects are used here; a
BYM2 decomposition can be substituted where a validated ABS adjacency
graph is available (\citeproc{ref-riebler2016}{Riebler et al. 2016}).

Because \(L_{rst}\) is standardised, \(\exp(\beta_L)\) is the
multiplicative change in expected rate per standard-deviation increase
in the lagged log count. For large counts the feedback behaves
approximately as \(Y_{t-1}^{\beta_L}\), so values near or above one
indicate near-explosive persistence and warrant posterior predictive
scrutiny.

\subsection{Baseline epidemic curve and intervention
effect}\label{baseline-epidemic-curve-and-intervention-effect}

The baseline curve is a Fourier expansion on the epidemic clock,

\[
f(\tau)=\sum_{h=1}^{H}\bigl\{a_h\sin(2\pi h\tau/52)+b_h\cos(2\pi h\tau/52)\bigr\},
\qquad
a_h,b_h\sim N(0,\sigma_f^2h^{-2p}).
\]

Two harmonics, as in a conventional seasonal regression, cannot
represent a sharply peaked epidemic with long shoulders. Rather than fixing
\(H\) small, \(H\) is taken to be moderate and higher harmonics are shrunk 
through the decay exponent \(p\), so the data determine peak sharpness while the
prior maintains smoothness; \(H=4\), \(p=1\) is the default and \(H=2\),
\(p=0\) recovers the two-harmonic model. Amplitude
\(R_h=(a_h^2+b_h^2)^{1/2}\) and phase \(\mathrm{atan2}(a_h,b_h)\) are
reported in preference to the individual coefficients.

The intervention acts multiplicatively with a within-season modulation
of free phase,

\[
m(\tau)=\beta_0+\beta_1\cos(2\pi\tau/52)+\beta_2\sin(2\pi\tau/52),
\qquad
RR(I,\tau)=\exp\{I\,m(\tau)\}.
\]

Here \(\beta_0\) is the cycle-average log rate ratio at reference
intensity \(I=1\), and \((\beta_1,\beta_2)\) describe how that effect
varies across the season. A single cosine term, as used in earlier
formulations, fixes the modulation phase at \(\tau=0\), which is an
arbitrary point in the epidemic: a modulation in quadrature with that
reference would return a null coefficient regardless of its magnitude. With
both terms free, the modulation amplitude
\((\beta_1^2+\beta_2^2)^{1/2}\) is reported. The intensity enters log-linearly; an
optional quadratic term allows this to be assessed rather than assumed.
When \(I=0\) or \(T_s=0\) the intervention contributes exactly zero.

\subsection{Identification}\label{identification}

Four sources of variation are separated: timing through
\((\delta_s,\kappa_s)\), season amplitude through \(v_s\), persistent
area risk through \(u_r\), and intervention-associated amplitude through
\(I_{rs}m\{\tau_s(t)\}\). The zero-sum constraints anchor the clock
exactly, so no residual location or scale drift is available to trade
against the Fourier phase.

Identification of \(\beta_0\) rests on between-area variation in
intensity within programme seasons. Season random effects absorb any
programme-season-wide amplitude change, by construction: that is what
makes the design robust to the concurrent secular changes which threaten
simple pre--post comparisons, and it is also what makes the estimate
uninformative if every area receives the same intensity. In the limit of
constant intensity across treated areas, \(I_{rs}m(\tau)\) becomes
collinear with \(v_s\) and \(\beta_0\) reverts to its prior. This is not
a failure mode to be discovered post hoc; it is a positivity condition
to be checked before fitting, and Section 3.2 reports the corresponding
diagnostics.

Uncertainty about \(\delta_s\) and \(\kappa_s\) enters the posterior for
\(\beta_0\), \(\beta_1\), \(\beta_2\) and every counterfactual estimand.
That propagation is the substantive difference from deterministic
registration followed by regression on the aligned scale, and the
simulation in Section 3.1 quantifies the impact of propagating this uncertainty.

\subsection{Priors and computation}\label{priors-and-computation}

All priors are proper, which with the negative binomial likelihood gives
a proper posterior. The default specification is

\[
\begin{aligned}
\alpha&\sim N(-6,2^2), &
\sigma_u,\sigma_v,\sigma_w&\sim N^+(0,0.5^2),\\
\sigma_\delta&\sim N^+(0,3^2), &
\sigma_\kappa&\sim N^+(0,0.15^2),\\
\sigma_f&\sim N^+(0,1), &
\beta_L&\sim N(0,0.5^2),\\
\beta_{\mathrm{SES}}&\sim N(0,0.3^2), &
\beta_0&\sim N(0,0.5^2),\\
\beta_1,\beta_2&\sim N(0,0.25^2), &
\phi^{-1}&\sim\mathrm{Exponential}(1).
\end{aligned}
\]

The shift prior places most mass within about six weeks of the average
clock and the speed prior favours modest dilation, both consistent with
observed inter-seasonal variation in RSV timing. The parameter \(\beta_0\) 
is deliberately centred at the null. Centring it on a protective effect, as would be
defensible given existing Nirsevimab evidence, makes the resulting
estimate difficult to distinguish from its prior; that specification is 
reported as a declared sensitivity analysis instead. The dispersion
is parameterised through \(\phi^{-1}\), which behaves better near the
Poisson limit than bounding \(\phi\) directly.

Inference uses Hamiltonian Monte Carlo through CmdStanR. Rank-normalised 
\(\widehat R\), bulk and tail effective sample size,
divergent transitions, energy diagnostics and tree-depth saturation are monitored, 
and fit is assessed with posterior predictive checks
(\citeproc{ref-gabry2019}{Gabry et al. 2019}). Predictive comparisons
between specifications use leave-one-season-out cross-validation rather
than pointwise leave-one-out. Pointwise leave-one-out is not appropriate
here: with a lagged outcome covariate, omitting a single week does not
remove its information, which survives in the predictor for the
following week, and the area--season effect induces further dependence
within blocks (\citeproc{ref-vehtari2017}{Vehtari, Gelman, and Gabry
2017}).

\subsection{Posterior g-computation}\label{posterior-g-computation}

Causal contrasts are obtained by the parametric g-formula
(\citeproc{ref-robins1986}{Robins 1986};
\citeproc{ref-hernanrobins2020}{Hernán and Robins 2020};
\citeproc{ref-snowden2011}{Snowden, Rose, and Mortimer 2011};
\citeproc{ref-keil2014}{Keil et al. 2014}). For each posterior draw
\(b\) all parameters are held at their sampled values and the expected
outcome is evaluated twice, under observed intensity and under zero
intensity, sharing the same clock, random effects, covariates and
offsets:

\[
C^{(b)}_{rs}=\sum_t\bigl\{\mu^{(b)}_{rst}(0)-\mu^{(b)}_{rst}(I_{rs})\bigr\},
\qquad
PF^{(b)}_{rs}=1-\frac{\sum_t\mu^{(b)}_{rst}(I_{rs})}{\sum_t\mu^{(b)}_{rst}(0)} .
\]

Summarising across draws propagates uncertainty in the coefficients, the
alignment, the random effects and the dispersion into the causal
contrasts.

The lagged-history term requires a choice. Evaluating both arms at the
observed lag gives a controlled contrast, in which
intensity changes while recent recorded history is held fixed. Because
\(\beta_L>0\) in these data, holding history fixed suppresses the onward
transmission consequences of prevented cases and attenuates the estimand
toward the null. A dynamic contrast is therefore also computed 
in which the history is propagated within each arm: within an
ABS--season, weeks are traversed in order and the lag entering week
\(t\) is reconstructed from a posterior predictive draw for week \(t-1\)
under that arm, so that both arms are simulated on the same
footing. Reporting both, and the gap between them, separates the direct
programme effect from its indirect transmission consequences under the
fitted dynamics. The dynamic contrast requires the autoregressive
specification to be interpreted as a transmission model, so the controlled 
contrast is treated as primary and the dynamic contrast as a bound on what 
the feedback term implies.

\subsection{Sensitivity to differential
ascertainment}\label{sensitivity-to-differential-ascertainment}

If a programme changes the propensity to consult or to code a syndromic
diagnosis, observed counts are distorted by a mechanism no temporal
alignment can recover. This is handled as a Bayesian bias analysis rather
than as a caveat. A multiplier \(\exp(-\lambda I_{rs})\) is applied to
programme-season observations in the likelihood but not in the
estimands. Since \(\lambda\) enters the likelihood only through
\(\beta_0-\lambda\), the data identify that difference and cannot
separate a programme effect from a programme-induced change in
ascertainment. The consequence is that prior uncertainty about
\(\lambda\) is added to the posterior for \(\beta_0\) rather than
assumed away, giving approximately
\(\operatorname{Var}(\beta_0)=\operatorname{Var}(\beta_0-\lambda\mid\text{data})+\sigma^2_\lambda\).
Setting \(\sigma_\lambda=0\) recovers the base model exactly. This
converts an untestable assumption into a reportable quantity: how large
would differential ascertainment have to be to overturn the conclusion?

\subsection{Simulation study}\label{simulation-study}

The simulation uses two tiers, combining precise Monte Carlo evaluation
of inexpensive estimators with a computationally feasible validation of
the full Bayesian model. The large-scale tier runs 500 repetitions per
scenario for the three comparators; the Bayesian tier runs 100
repetitions per scenario with the actual Stan program.

Each dataset has 10 areas, six seasons and 52 weekly observations per
season, with the programme introduced in seasons 5 and 6. Area
populations are log-normal with median approximately 950 infants and
log-scale standard deviation 0.22, truncated below at 300. Area and
season log-rate effects have standard deviations 0.25 and 0.15; a
standardised area-level socioeconomic covariate has coefficient 0.15;
the lag covariate is \(\ell_{rst}=\{\log(1+Y_{rs,t-1})-1\}/1.25\) with
coefficient 0.10. Counts are generated as
\(Y_{rst}\sim\mathrm{NB}_2(\mu_{rst},\phi)\) with \(\phi=12\) and

\[
\begin{aligned}
\log\mu_{rst}={}&\log N_{rst}+\alpha+u_r+v_s+f(\tau_{st})+\beta_{\mathrm{lag}}\ell_{rst}
+\beta_{\mathrm{SES}}\mathrm{SES}_r\\
&+\mathbb{I}(s\geq5)I_r\beta_I+\xi_{rs},
\end{aligned}
\]

where \(\alpha=-6.35\), \(\beta_I=\log(0.55)\),
\(\tau_{st}=(t-\delta_s)\exp(\kappa_s)\), \(\xi_{rs}\sim N(0,0.15^2)\)
is an area--season disturbance held constant across the 52 weeks of a
block, that is, at exactly the level at which programme intensity
varies, and

\[
f(\tau)=0.75\sin(2\pi\tau/52)-0.15\cos(2\pi\tau/52)+0.30\sin(4\pi\tau/52)-0.10\cos(4\pi\tau/52).
\]

The intervention coefficient corresponds to a 45\% rate reduction at
reference intensity. The within-season modulation is set to zero so that
false-positive behaviour and shrinkage can be assessed. Programme
intensity is generated as
\(I^*_r=\exp\{0.20\,\mathrm{SES}_r+cU_r+\epsilon_r\}\) with
\(\epsilon_r\sim N(0,0.18^2)\), normalised by its cross-area median,
where \(U_r\sim N(0,1)\) is an unobserved access variable and \(c\)
governs intensity--risk confounding.

The four scenarios are:

\begin{enumerate}
\def\labelenumi{\arabic{enumi}.}
\tightlist
\item
  Aligned. \(\delta_s=\kappa_s=0\); no ascertainment change or
  confounding.
\item
  Phase shift. Season-specific displacement and speed
  variation, with raw standard deviations 2.5 weeks and 0.08 on the
  log-speed scale, centred across seasons and bounded to \([-6,6]\)
  weeks and \([-0.25,0.25]\).
\item
  Phase and ascertainment. As (2), plus intensity-dependent
  post-programme ascertainment through the multiplier
  \(\exp(-0.15I_r)\).
\item
  Phase and confounding. As (2), with \(c=0.60\) and an
  additional programme-period outcome term
  \(0.132U_r\mathbb{I}(s\geq5)\).
\end{enumerate}

Scenarios 3 and 4 deliberately violate the fitted outcome model. Their
role is to mark the boundary between bias arising from epidemic-clock
variation, which alignment addresses, and bias arising from the
observation or treatment-assignment mechanisms, which it cannot.

Three inexpensive comparators are evaluated. The interrupted
time-series (ITS) estimator follows the segmented-regression principle
(\citeproc{ref-bernal2017}{Bernal, Cummins, and Gasparrini 2017}): a
negative binomial regression indexed by calendar time with a
six-degree-of-freedom natural spline for week, fixed area and season
effects, the lag and socioeconomic covariates and an
intensity-by-programme term. The generalized additive model
(GAM) replaces that spline by a penalised cyclic smooth of basis
dimension 12 with random-effect smooths for area and season
(\citeproc{ref-wood2017}{Wood 2017}). Both remain indexed by calendar
week. The two-stage aligned negative binomial (Aligned NB)
applies deterministic registration (\citeproc{ref-ramsay1998}{Ramsay and
Li 1998}): a pre-programme template is built from area-aggregated
incidence, season-specific shifts in \([-5,5]\) weeks and speeds in
\([0.90,1.10]\) are chosen to maximise correlation with the interpolated
template, and a two-harmonic negative binomial regression is fitted on
the resulting fixed clock. Because the alignment is treated as known at
the second stage, its uncertainty does not reach the interval for the
intervention effect. Contrasting these three with the proposed model
isolates the consequences of calendar indexing, deterministic
preprocessing and joint Bayesian alignment respectively.

Each Bayesian fit in the simulation tier uses four chains with 300
warm-up and 300 retained iterations, step size 0.90 and
maximum tree depth 11, from dispersed random initial values. These are
short chains, adopted so that the full Bayesian tier remains feasible;
initialising near the data-generating values would make a coverage study
circular, and is not used. As a check that the abbreviated schedule does
not itself distort calibration, a random subset of fits per scenario is
repeated with 1000 warm-up and 1000 retained iterations and the
operating characteristics compared. The empirical analysis uses the
longer schedule with step size 0.95. Performance is
summarised by Monte Carlo bias and root mean squared error relative to
\(\beta_I\), empirical coverage and mean width of nominal 95\% intervals
(Wald for the comparators, equal-tailed posterior for the Bayesian
model), with Monte Carlo standard errors for bias and coverage. The
Bayesian tier additionally records \(\widehat R\), bulk effective sample
size, divergences, tree-depth saturation, recovery of \(\phi\) and root
mean squared errors for the season-specific shift and log-speed
parameters.

\subsection{Catalan open data}
Two open datasets from the Generalitat de Catalunya are combined. The Information System for Surveillance of Infections in Catalonia (SIVIC) primary-care surveillance file holds daily counts by date, ABS, diagnosis, age group, sex, socioeconomic index and population denominator; the immunisation file holds campaign-specific numbers immunised and reference populations by ABS and demographic stratum.

Records with a diagnosis containing the Catalan or Spanish stem for bronchiolitis and an infant age category were selected, aggregated to Monday-based epidemiological weeks and July-to-June seasons. Within each ABS-week, cases were summed, the population denominator retained and the socioeconomic index population-weighted. RSV immunisation records were identified from campaign labels containing \textit{VRS}, \textit{respiratory syncytial virus} or \textit{Nirsevimab}.

The immunisation file reports campaign-accumulated counts against a cross-sectional reference population. Because the infant cohort is open during the campaign, the accumulated count can exceed that population and the ratio is not a coverage estimate. Raw intensity is defined as $I^{\mathrm{raw}}_{rs}=M_{rs}/P^{\mathrm{ref}}_{rs}$ and scaled by the median over programme ABS--seasons,

\[
I_{rs}=I^{\mathrm{raw}}_{rs}\big/\operatorname{median}_{r,s:T_s=1}\bigl(I^{\mathrm{raw}}_{rs}\bigr),
\]

so that $I_{rs}=0$ denotes no recorded activity and $I_{rs}=1$ the median programme intensity. This is an ecological measure of programme activity, not individual uptake.

Descriptive analysis used every ABS with a valid weekly denominator. The model analysis retained weeks 1--52 of complete seasons and the 60 ABS with the largest infant populations, classifying seasons beginning in 2023 or later as programme seasons. Pre-programme seasons identify the common epidemic curve and the distribution of timing parameters; area-specific intensity variation within programme seasons identifies the intervention contrast. Pandemic-era deviations are absorbed by unrestricted season effects.

\section{Results}\label{results}

\subsection{Simulation}\label{simulation}

\begin{table}
\centering
\caption{\label{tab:simulation-table}Operating characteristics for the log rate ratio, based on 500 repetitions per scenario for the comparators and 100 for the proposed model. Fits reports successful/requested; Monte Carlo standard errors in parentheses.}
\centering
\resizebox{\ifdim\width>\linewidth\linewidth\else\width\fi}{!}{
\fontsize{7}{9}\selectfont
\begin{tabular}[t]{>{\raggedright\arraybackslash}p{2.6cm}>{\raggedright\arraybackslash}p{2.2cm}lllll}
\toprule
Scenario & Method & Fits & Bias (MCSE) & RMSE & Coverage, \% (MCSE) & Width\\
\midrule
Aligned & Aligned NB & 500/500 & 0.023 (0.010) & 0.222 & 82.8 (1.7) & 0.598\\
Aligned & BEACON-Stan & 100/100 & 0.044 (0.011) & 0.121 & 93.0 (2.6) & 0.503\\
Aligned & GAM & 500/500 & 0.056 (0.009) & 0.198 & 35.4 (2.1) & 0.194\\
Aligned & ITS & 500/500 & 0.022 (0.010) & 0.222 & 83.0 (1.7) & 0.598\\
\addlinespace
Phase + ascertainment & Aligned NB & 500/500 & -0.117 (0.010) & 0.258 & 77.6 (1.9) & 0.623\\
Phase + ascertainment & BEACON-Stan & 100/100 & -0.117 (0.012) & 0.170 & 87.0 (3.4) & 0.545\\
Phase + ascertainment & GAM & 500/500 & -0.037 (0.008) & 0.186 & 40.2 (2.2) & 0.203\\
Phase + ascertainment & ITS & 500/500 & -0.079 (0.010) & 0.235 & 82.4 (1.7) & 0.634\\
\addlinespace
Phase + confounding & Aligned NB & 500/500 & 0.172 (0.005) & 0.200 & 30.2 (2.1) & 0.254\\
Phase + confounding & BEACON-Stan & 100/100 & 0.113 (0.009) & 0.144 & 68.0 (4.7) & 0.333\\
Phase + confounding & GAM & 500/500 & 0.145 (0.006) & 0.193 & 29.6 (2.0) & 0.155\\
Phase + confounding & ITS & 500/500 & 0.190 (0.004) & 0.215 & 24.2 (1.9) & 0.258\\
\addlinespace
Phase shift & Aligned NB & 500/500 & 0.018 (0.011) & 0.236 & 79.2 (1.8) & 0.592\\
Phase shift & BEACON-Stan & 100/100 & 0.052 (0.013) & 0.140 & 89.0 (3.1) & 0.513\\
Phase shift & GAM & 500/500 & 0.094 (0.008) & 0.194 & 35.0 (2.1) & 0.194\\
Phase shift & ITS & 500/500 & 0.051 (0.010) & 0.229 & 79.6 (1.8) & 0.604\\
\bottomrule
\end{tabular}}
\end{table}

In the aligned scenario, estimated bias was ITS 0.022, GAM 0.056,
Aligned NB 0.023, and BEACON-Stan 0.044. In the phase-shift scenario,
estimated bias was ITS 0.051, GAM 0.094, Aligned NB 0.018, and
BEACON-Stan 0.052. Empirical coverage of the nominal 95\% credible
interval was 93.0\% under stable timing and 89.0\% under epidemic-clock
variation. Under the two deliberately misspecified scenarios, bias was
-0.117 with intensity-dependent ascertainment and 0.113 with unmeasured
intensity--risk confounding, and coverage fell to 87.0\% and 68.0\%
respectively. These figures quantify the residual bias that temporal
alignment alone cannot remove.

\begin{table}
\centering
\caption{\label{tab:beacon-recovery-table}Parameter recovery and sampling diagnostics. Values in parentheses are root mean squared errors; timing errors are reported for the season-specific shift and log speed. The final column gives total divergent transitions and tree-depth saturations.}
\centering
\resizebox{\ifdim\width>\linewidth\linewidth\else\width\fi}{!}{
\fontsize{7}{9}\selectfont
\begin{tabular}[t]{>{\raggedright\arraybackslash}p{2.6cm}rlllllrl}
\toprule
Scenario & Fits & Intensity bias (RMSE) & Modulation bias & Phi bias & Timing RMSE: shift / speed & R-hat: mean / max & Bulk ESS & Div. / depth\\
\midrule
Aligned & 100 & 0.044 (0.121) & 0.003 & 0.73 & 0.16 / 0.007 & 1.006 / 1.035 & 571 & 134 / 0\\
Phase + ascertainment & 100 & -0.117 (0.170) & -0.004 & 0.30 & 1.70 / 0.030 & 1.005 / 1.018 & 640 & 21 / 0\\
Phase + confounding & 100 & 0.113 (0.144) & 0.008 & 0.91 & 1.77 / 0.028 & 1.004 / 1.019 & 736 & 15 / 0\\
Phase shift & 100 & 0.052 (0.140) & -0.001 & 0.60 & 1.69 / 0.029 & 1.005 / 1.016 & 602 & 29 / 0\\
\bottomrule
\end{tabular}}
\end{table}

\begin{figure}[H]

\centering{

\pandocbounded{\includegraphics[keepaspectratio]{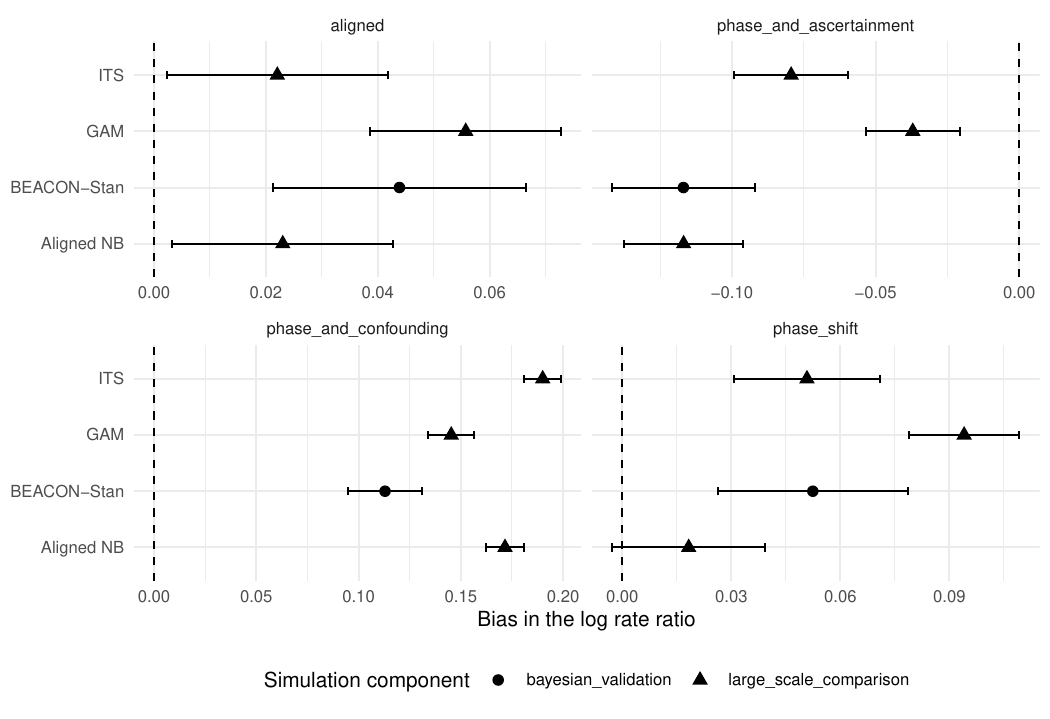}}

}

\caption{\label{fig-simulation-bias}Bias in the log rate ratio across
simulation scenarios. Horizontal segments span two Monte Carlo standard
errors.}

\end{figure}%

The two tiers answer different questions. The large-scale tier
characterises the inexpensive calendar-time and deterministic-alignment
alternatives with small Monte Carlo error; the Bayesian tier validates
the model actually used in the application and confirms that timing
uncertainty reaches the intervention posterior. Within that design, the
aligned scenario measures the cost of estimating a clock that is not
needed, and the phase-shift scenario measures the benefit when it is.
Scenarios 3 and 4 bound the benefit: both perturb observed amplitude
through mechanisms that temporal registration does not address, and no
estimator that conditions only on the modelled covariates should be
expected to recover the truth under them.

\subsection{Catalan data and overlap}\label{catalan-data-and-overlap}

The analytic series covered 03 October 2011 to 29 June 2026, spanning 15
epidemic seasons and 377 basic health areas, with 178,355 infant
bronchiolitis diagnoses recorded. The population-weighted mean scaled
RSV immunisation intensity across programme seasons was 0.98.

\begin{longtable}[]{@{}
  >{\raggedright\arraybackslash}p{(\linewidth - 10\tabcolsep) * \real{0.0847}}
  >{\raggedright\arraybackslash}p{(\linewidth - 10\tabcolsep) * \real{0.1017}}
  >{\raggedleft\arraybackslash}p{(\linewidth - 10\tabcolsep) * \real{0.2966}}
  >{\raggedright\arraybackslash}p{(\linewidth - 10\tabcolsep) * \real{0.0932}}
  >{\raggedleft\arraybackslash}p{(\linewidth - 10\tabcolsep) * \real{0.1864}}
  >{\raggedright\arraybackslash}p{(\linewidth - 10\tabcolsep) * \real{0.2373}}@{}}
\caption{Infant bronchiolitis and scaled RSV immunisation intensity by
epidemic season.}\tabularnewline
\toprule\noalign{}
\begin{minipage}[b]{\linewidth}\raggedright
Season
\end{minipage} & \begin{minipage}[b]{\linewidth}\raggedright
Total cases
\end{minipage} & \begin{minipage}[b]{\linewidth}\raggedleft
Incidence per 100,000 person-weeks
\end{minipage} & \begin{minipage}[b]{\linewidth}\raggedright
Peak week
\end{minipage} & \begin{minipage}[b]{\linewidth}\raggedleft
Peak weekly incidence
\end{minipage} & \begin{minipage}[b]{\linewidth}\raggedright
Mean immunisation intensity
\end{minipage} \\
\midrule\noalign{}
\endfirsthead
\toprule\noalign{}
\begin{minipage}[b]{\linewidth}\raggedright
Season
\end{minipage} & \begin{minipage}[b]{\linewidth}\raggedright
Total cases
\end{minipage} & \begin{minipage}[b]{\linewidth}\raggedleft
Incidence per 100,000 person-weeks
\end{minipage} & \begin{minipage}[b]{\linewidth}\raggedright
Peak week
\end{minipage} & \begin{minipage}[b]{\linewidth}\raggedleft
Peak weekly incidence
\end{minipage} & \begin{minipage}[b]{\linewidth}\raggedright
Mean immunisation intensity
\end{minipage} \\
\midrule\noalign{}
\endhead
\bottomrule\noalign{}
\endlastfoot
2011-2012 & 11,636 & 1654.9 & 2011-12-12 & 2981.7 & -- \\
2012-2013 & 13,252 & 1696.4 & 2012-12-17 & 2966.8 & -- \\
2013-2014 & 12,884 & 1707.2 & 2013-12-30 & 3220.6 & -- \\
2014-2015 & 14,066 & 1852.9 & 2014-12-29 & 3252.5 & -- \\
2015-2016 & 14,936 & 2068.2 & 2015-12-28 & 4369.9 & -- \\
2016-2017 & 14,465 & 2004.3 & 2016-12-12 & 3801.5 & -- \\
2017-2018 & 14,145 & 1932.3 & 2017-12-11 & 3641.2 & -- \\
2018-2019 & 14,455 & 2063.2 & 2018-12-17 & 3610.7 & -- \\
2019-2020 & 11,126 & 2240.2 & 2019-12-30 & 3902.5 & -- \\
2020-2021 & 4,310 & 1576.4 & 2021-06-21 & 2038.8 & -- \\
2021-2022 & 11,140 & 2014.9 & 2021-11-22 & 3127.5 & -- \\
2022-2023 & 15,131 & 2364.5 & 2022-11-21 & 4771.0 & -- \\
2023-2024 & 10,442 & 1860.5 & 2023-12-11 & 2828.6 & 0.9 \\
2024-2025 & 8,371 & 1671.7 & 2024-12-30 & 2043.3 & 1.01 \\
2025-2026 & 7,996 & 1625.0 & 2025-12-29 & 2002.7 & 1.02 \\
\end{longtable}

\begin{figure}[H]

\centering{

\pandocbounded{\includegraphics[keepaspectratio]{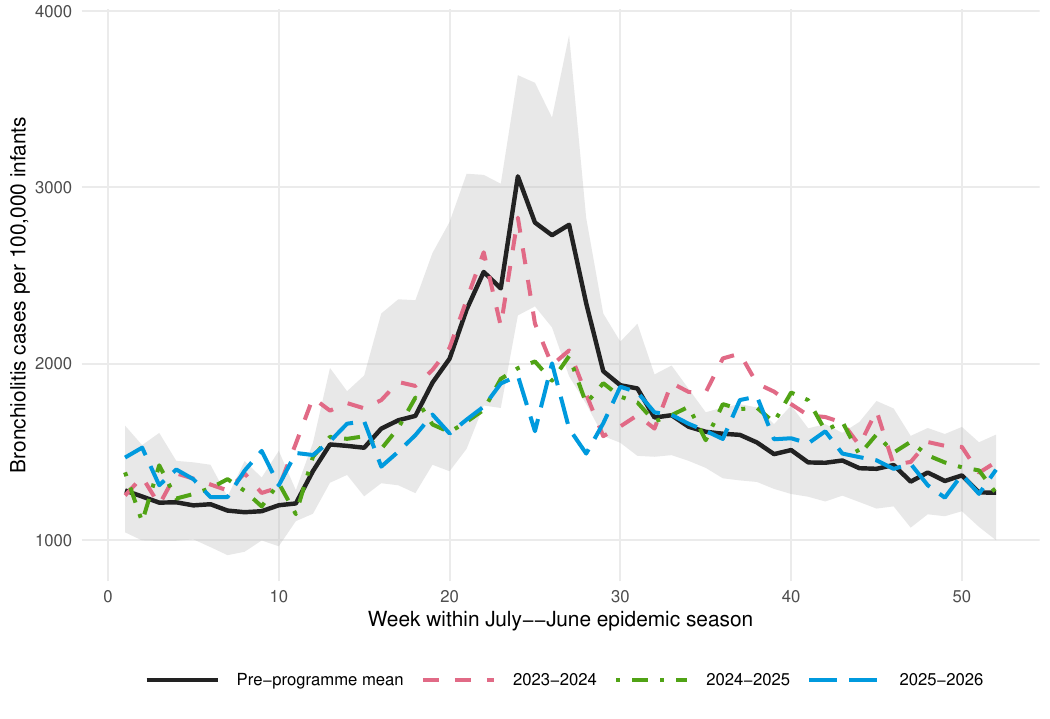}}

}

\caption{\label{fig-epidemic-profile}Weekly infant bronchiolitis
incidence in programme seasons against the pre-programme distribution.
The ribbon spans the 10th--90th percentile across pre-programme
seasons.}

\end{figure}%

\begin{figure}[H]

\centering{

\pandocbounded{\includegraphics[keepaspectratio]{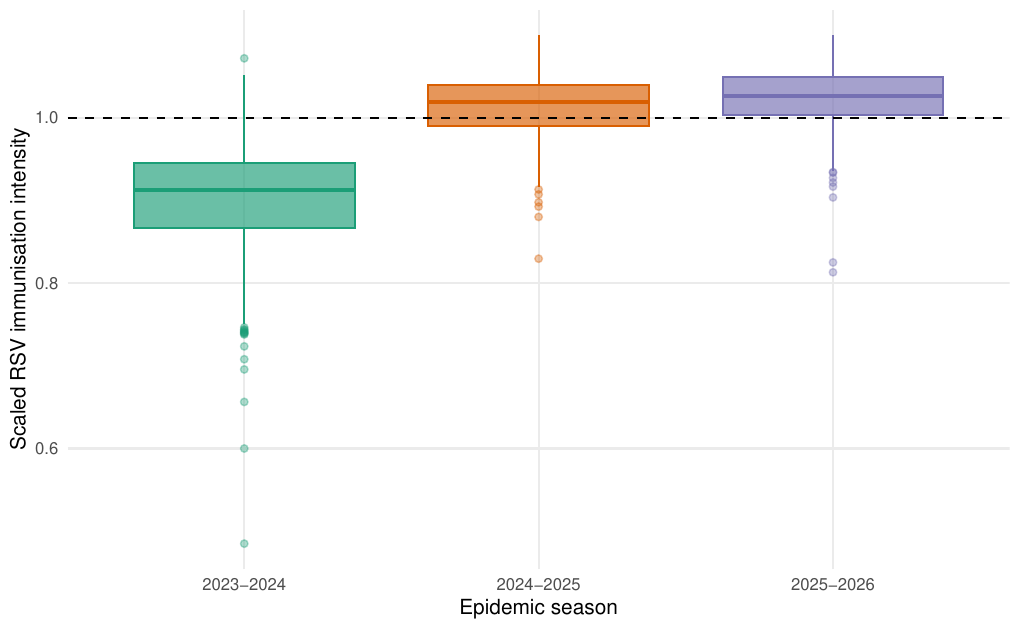}}

}

\caption{\label{fig-intensity}Distribution of scaled RSV immunisation
intensity across basic health areas by programme season. The spread
within a season, not its level, is what identifies the intervention
effect.}

\end{figure}%

Because the intervention contrast is identified by between-area
variation in intensity within programme seasons, the dispersion visible
in Figure~\ref{fig-intensity}, rather than its level, determines how
much the data can say about \(\beta_0\). Overlap is therefore reported 
explicitly before any effect estimate.

\begin{longtable}[]{@{}
  >{\raggedright\arraybackslash}p{(\linewidth - 12\tabcolsep) * \real{0.1389}}
  >{\raggedleft\arraybackslash}p{(\linewidth - 12\tabcolsep) * \real{0.0556}}
  >{\raggedright\arraybackslash}p{(\linewidth - 12\tabcolsep) * \real{0.0972}}
  >{\raggedright\arraybackslash}p{(\linewidth - 12\tabcolsep) * \real{0.0833}}
  >{\raggedright\arraybackslash}p{(\linewidth - 12\tabcolsep) * \real{0.0833}}
  >{\raggedright\arraybackslash}p{(\linewidth - 12\tabcolsep) * \real{0.2639}}
  >{\raggedright\arraybackslash}p{(\linewidth - 12\tabcolsep) * \real{0.2778}}@{}}
\caption{Overlap in scaled immunisation intensity across basic health
areas within each programme season. A high proportion of areas close to
the season median indicates a near-degenerate dose-response
contrast.}\tabularnewline
\toprule\noalign{}
\begin{minipage}[b]{\linewidth}\raggedright
Season
\end{minipage} & \begin{minipage}[b]{\linewidth}\raggedleft
ABS
\end{minipage} & \begin{minipage}[b]{\linewidth}\raggedright
Median
\end{minipage} & \begin{minipage}[b]{\linewidth}\raggedright
SD
\end{minipage} & \begin{minipage}[b]{\linewidth}\raggedright
IQR
\end{minipage} & \begin{minipage}[b]{\linewidth}\raggedright
10th--90th centile
\end{minipage} & \begin{minipage}[b]{\linewidth}\raggedright
Within 5\% of median
\end{minipage} \\
\midrule\noalign{}
\endfirsthead
\toprule\noalign{}
\begin{minipage}[b]{\linewidth}\raggedright
Season
\end{minipage} & \begin{minipage}[b]{\linewidth}\raggedleft
ABS
\end{minipage} & \begin{minipage}[b]{\linewidth}\raggedright
Median
\end{minipage} & \begin{minipage}[b]{\linewidth}\raggedright
SD
\end{minipage} & \begin{minipage}[b]{\linewidth}\raggedright
IQR
\end{minipage} & \begin{minipage}[b]{\linewidth}\raggedright
10th--90th centile
\end{minipage} & \begin{minipage}[b]{\linewidth}\raggedright
Within 5\% of median
\end{minipage} \\
\midrule\noalign{}
\endhead
\bottomrule\noalign{}
\endlastfoot
2023-2024 & 60 & 0.92 & 0.049 & 0.054 & 0.87--0.96 & 75\% \\
2024-2025 & 60 & 1.02 & 0.035 & 0.052 & 0.97--1.05 & 87\% \\
2025-2026 & 60 & 1.03 & 0.030 & 0.037 & 0.99--1.06 & 93\% \\
\end{longtable}

Overlap differs markedly between programme seasons. In 2023-2024 the
interdecile range of scaled intensity was 0.87 to 0.96, whereas in
2025-2026 93\% of areas lay within 5\% of the season median. The
dose-response contrast is therefore informed disproportionately by the
earlier campaign, and the posterior for the intervention coefficient
should be read with that in mind.

\subsection{Posterior intervention
estimates}\label{posterior-intervention-estimates}

The posterior cycle-average rate ratio at the reference immunisation
intensity was 1.01 (95\% CrI 0.83 to 1.20). Under the controlled
contrast, the posterior median prevented fraction was 5.3\% (95\% CrI
-12.6\% to 21.2\%), corresponding to 411 (95\% CrI -817 to 1971)
primary-care bronchiolitis events across the analysed programme seasons.
Propagating disease history within each arm gives a dynamic prevented
fraction of 6.8\% (95\% CrI -13.9\% to 25.1\%); the difference between
the two contrasts measures the onward transmission effect implied by the
fitted autoregressive term. The estimated area--season standard
deviation was 0.04 (95\% CrI 0.01 to 0.06), confirming residual
clustering at the level at which the exposure varies.

\subsection{External benchmarks}\label{external-benchmarks}

Published Catalan analyses provide external reference points, though
none estimates the same quantity. A 2023--2024 cohort study reported
reductions of approximately 87.6\% in RSV-bronchiolitis hospital
admission and 90.1\% in intensive-care admission among immunised infants
(\citeproc{ref-coma2024}{Coma et al. 2024}). A population-level analysis
found the relative burden of all-cause bronchiolitis among infants aged
0--11 months fell by about 40\% relative to older infants after
programme introduction (\citeproc{ref-perramon2024}{Perramon-Malavez et
al. 2024}). For the 2024--2025 campaign, effectiveness against
primary-care RSV-related infection was 76.1\% (60.3\% to 85.6\%)
(\citeproc{ref-coma2025}{Coma et al. 2025}).

These are individual-level effectiveness estimates among immunised
infants, conditional on eligibility and, in the cohort studies, on a
pathogen-confirmed outcome. The estimand here is a population-level
contrast in syndromic burden between observed and zero programme
intensity, across all infants in an area regardless of immunisation
status, identified by between-area contrasts in programme activity. The
two need not agree even if both are correct: high individual
effectiveness is compatible with a modest population-level syndromic
contrast when the syndromic outcome includes substantial non-RSV
aetiology, when uptake is high and near-uniform so that the between-area
contrast is small, and when season effects absorb the programme-wide
component of any change. The comparison is therefore a check on
plausibility rather than a validation.

\section{Discussion}\label{discussion}

Calendar alignment embeds an assumption that seasonal epidemics recur on a fixed schedule. Since epidemics naturally vary in their timing, this assumption converts a shift in epidemic phase into an apparent change in disease
burden. Estimating the epidemic clock jointly with the intervention
effect removes that particular error, and does so adaptively: when the
data support little displacement the timing parameters shrink towards a
common calendar and the model behaves like a conventional seasonal count
regression; when displacement is supported, its uncertainty widens the
intervention posterior instead of being discarded by a deterministic
preprocessing step.

The simulation delimits what the proposed framework can address. Under stable timing, estimating
a clock that is not needed costs little. Under phase and speed
variation, joint alignment reduces bias and improves interval
calibration relative to both calendar-indexed regression and two-stage
registration. Under intensity-dependent ascertainment and unmeasured
intensity--risk confounding, coverage degrades for every method
considered, because both mechanisms alter observed amplitude through
channels absent from the fitted outcome model. Posterior credible
intervals are calibrated conditional on that model; they do not, and
cannot, absorb bias from an omitted observation or assignment process.
The apparently better coverage of the interrupted time-series estimator
under differential ascertainment should be interpreted with caution: its intervals are
substantially wider, and part of the misspecification is absorbed by the
calendar spline and season effects. The useful conclusion is a boundary.
Epidemic alignment addresses phase-induced bias; it does not address
measurement or confounding bias, and combining it with the ascertainment
sensitivity analysis described in Section 2.8 is the appropriate
response.

For the Catalan application, the overlap diagnostics are as important as
the effect estimate. The design deliberately relies on within-season
contrasts between areas so that season effects can absorb programme-wide
secular change, which is what protects against the concurrent-trend
threats that undermine simple pre--post comparisons. The cost is that
the estimate becomes uninformative when programme intensity is
near-uniform across areas, and in the later campaigns it largely is. A
wide posterior interval spanning the null is, under those circumstances,
the correct output of a well-behaved model rather than evidence about
Nirsevimab: it reports that these data, at this level of aggregation,
cannot separate a programme effect from a season effect. This distinction is crucial, because the alternative interpretation, that a highly
effective immunisation programme produced no detectable population
signal, is not supported by the design. The contrast with published
individual-level effectiveness estimates is consistent with this
interpretation, and the seasons with wider intensity dispersion are
where the informative comparison lies.

The framework yields estimands that map onto decisions rather than onto
coefficients: prevented events and fractions, peak attenuation,
displacement of the epidemic peak, and stratified contrasts by geography
and socioeconomic composition. Distinguishing the controlled from the
dynamic contrast matters here. Holding observed history fixed suppresses
the onward transmission consequences of prevented cases and attenuates
the estimand; propagating history within each arm restores them, at the cost of interpreting the autoregressive term strictly as a transmission mechanism. Reporting both bounds the sensitivity of the conclusion to
that choice. Separately, the estimated shift and speed distributions are
of epidemiological interest independent of any intervention, since they
quantify whether particular seasons were early, late, compressed or
prolonged relative to the historical clock.

The main structural limitation is that affine warping cannot represent
asymmetric deformation, in which growth and decline change differently;
monotone spline warps would relax this at the cost of a harder
identification problem. Bronchiolitis is an imperfect proxy for RSV, and
secular change in consultation or coding alters the observation process
in ways the ascertainment sensitivity parameter can bound but not
identify. Area-level intensity derived from accumulated campaign counts
against a cross-sectional reference population is an ecological measure
of programme activity, not coverage, and does not identify individual
immunisation status. Residual confounding persists to the extent that
unmeasured determinants of programme intensity also affect bronchiolitis
risk after conditioning on area, season, area-season, socioeconomic and
temporal terms; the simulation shows this is the failure mode against
which alignment offers no protection at all.

Despite these limitations, the primary conclusion remains robust: Epidemic alignment is a component
of the data-generating process, not a presentational convenience.
Modelling it explicitly separates temporal displacement from change in
disease amplitude and produces coherent uncertainty for the intervention
effect, and the same construction applies to vaccination programmes,
diagnostic-policy change, non-pharmaceutical interventions and
environmental perturbations affecting seasonal transmission.

\textbf{Funding.} This work is supported by grant PID2025-173749OB-I00 from the Spanish Ministry of
Science and Innovation and the Spanish State Research Agency, and the
Severo Ochoa and María de Maeztu Programme for Centres and Units of
Excellence in R\&D (CEX2020-001084-M).

\textbf{Data availability.} Both datasets are openly published by the
Generalitat de Catalunya through the Catalonia Open Data portal:
primary-care syndromic surveillance and
persons immunised by respiratory-virus campaign. No individual-level data were accessed.

\textbf{Code availability.} The Stan model, simulation code and analysis
scripts are available at \url{https://github.com/dmorinya/BEACON}.

\section*{References}\label{references}
\addcontentsline{toc}{section}{References}

\phantomsection\label{refs}

\newpage

\section*{Appendix A. Basic health areas in
Catalonia}\label{appendix-a.-basic-health-areas-in-catalonia}
\addcontentsline{toc}{section}{Appendix A. Basic health areas in
Catalonia}

The Catalan health system is organised into health regions, health
sectors and \emph{àrees bàsiques de salut} (ABSs). The ABS is the
smallest of these planning units and the level at which primary
healthcare is organised, each with a reference population and a primary
care team of general practitioners, paediatricians, nurses, social
workers and administrative staff. The term \emph{basic health area} is 
used throughout for the ABS and \emph{health region} is reserved for the 
larger unit.

ABS boundaries follow the organisation and accessibility of primary-care
services rather than general administrative divisions. An ABS may
coincide with a municipality, combine several in sparsely populated
areas, or subdivide a large municipality. The resulting pattern is
strongly heterogeneous: metropolitan Barcelona and other dense urban
corridors contain many small ABSs, while the Pyrenees, inland Catalonia
and the Terres de l'Ebre are covered by substantially larger
territories. ABS boundaries need not coincide with municipal, county or
provincial boundaries.

\begin{figure}[H]

\centering{

\includegraphics[width=0.88\linewidth,height=\textheight,keepaspectratio]{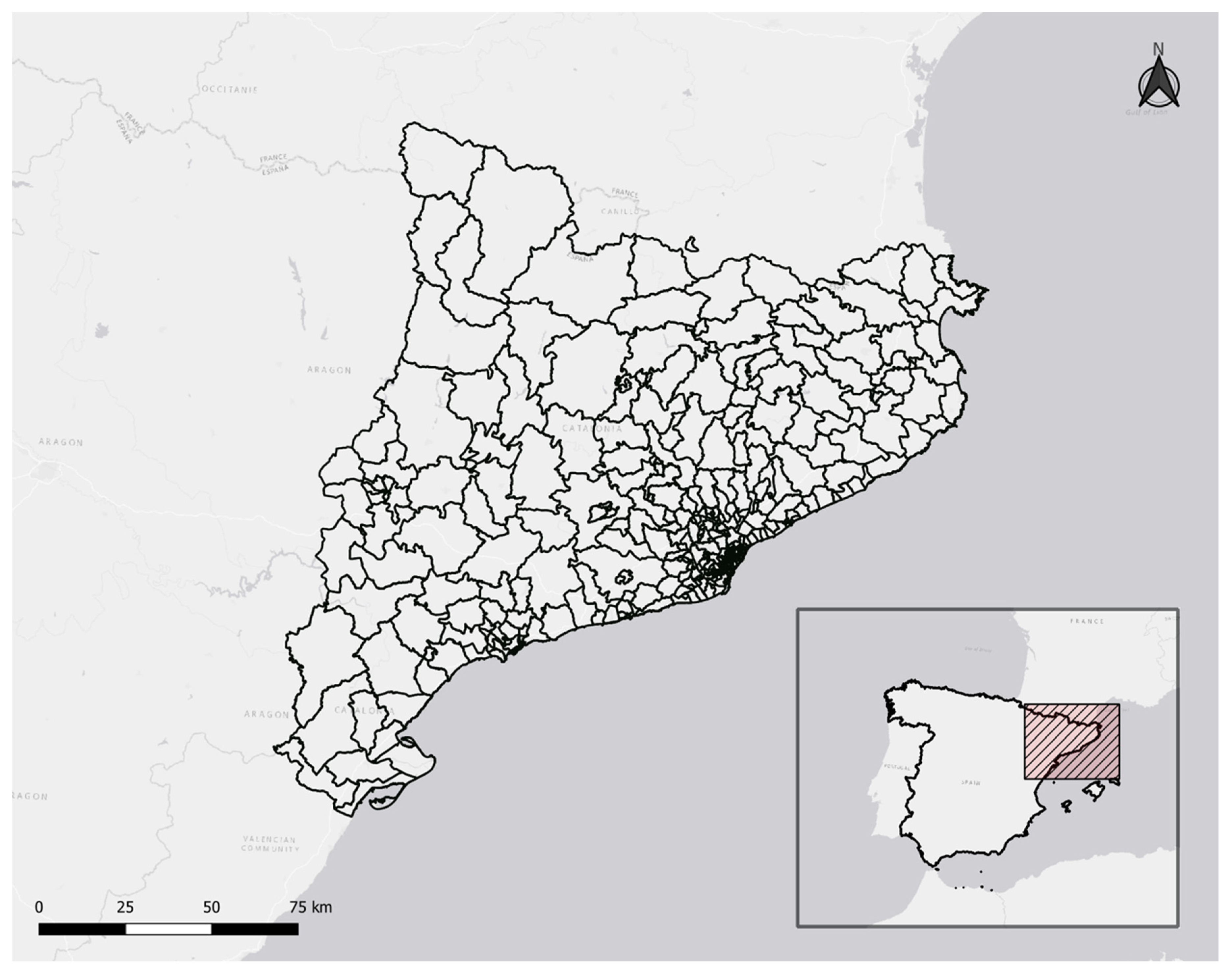}

}

\caption{\label{fig-abs-map}Geographical distribution of basic health
areas in Catalonia, showing the concentration of small areas in
metropolitan Barcelona against the larger rural and mountain
territories. Source: Chaudhuri et al.~(2022), reproduced under CC BY
4.0.}

\end{figure}

Figure~\ref{fig-abs-map} illustrates this distribution. The
configuration shown contains 343 ABSs and reflects cartography
documented in 2022; planning boundaries are revised over time, and the
surveillance extracts used here contain a larger number of ABS codes, so
the map is indicative of the spatial pattern rather than an exact key to
the analysis units. In the analysis, the ABS code supplied in the Information System for Surveillance of Infections in Catalonia (SIVIC)
and immunisation datasets links outcomes, population denominators,
socioeconomic information and programme intensity; an ABS--season is the
pairing of one ABS with one July--June epidemic season.

The map is reproduced from Chaudhuri, Giménez-Adsuar, Saez and Barceló
(2022), ``PandemonCAT: Monitoring the COVID-19 Pandemic in Catalonia,
Spain'', \emph{International Journal of Environmental Research and
Public Health}, 19, 4783
(\href{https://doi.org/10.3390/ijerph19084783}{doi:10.3390/ijerph19084783}),
which describes the ABS as the lowest health-related geographical
aggregation used for official small-area surveillance in Catalonia and
provides the digitised cartography.

\end{document}